\documentclass[aps,prl,reprint,twocolumn,twoside,showpacs,floatfix,letterpaper,superscriptaddress,bibnotes,footinbib]{revtex4-1}

\usepackage{graphicx}
\usepackage{dcolumn}
\usepackage{bm}
\usepackage{natbib}
\usepackage{multirow}
\usepackage{amsmath}
\usepackage{amssymb}
\usepackage{mathrsfs}
\usepackage{bbold}
\usepackage{hyperref}
\usepackage{aas_macros_prl}
\usepackage{color}

\newcommand{\hMpc}{h^{-1}{\rm Mpc}}
\newcommand{\hGpc}{h^{-1}{\rm Gpc}}
\newcommand{\hMsun}{h^{-1}M_{\odot}}
\newcommand{\void}{\mathrm{v}}

\newcommand{\gal}{\mathrm{g}}
\newcommand{\matter}{\mathrm{m}}
\begin{document}

\title{Cosmology with Void-Galaxy Correlations}

\author{Nico Hamaus}
\email{hamaus@iap.fr}
\affiliation{Institut d'Astrophysique de Paris, UMR 7095, CNRS - Universit\'e Pierre et Marie Curie (Univ Paris 06), 75014 Paris, France}
\affiliation{Department of Physics, University of Illinois at Urbana-Champaign, Urbana, IL 61801, USA}

\author{Benjamin D. Wandelt}
\affiliation{Institut d'Astrophysique de Paris, UMR 7095, CNRS - Universit\'e Pierre et Marie Curie (Univ Paris 06), 75014 Paris, France}
\affiliation{Department of Physics, University of Illinois at Urbana-Champaign, Urbana, IL 61801, USA}

\author{P. M. Sutter}
\affiliation{Institut d'Astrophysique de Paris, UMR 7095, CNRS - Universit\'e Pierre et Marie Curie (Univ Paris 06), 75014 Paris, France}
\affiliation{Department of Physics, University of Illinois at Urbana-Champaign, Urbana, IL 61801, USA}
\affiliation{Center for Cosmology \& AstroParticle Physics, Ohio State University, Columbus, OH 43210, USA}

\author{Guilhem Lavaux}
\affiliation{Institut d'Astrophysique de Paris, UMR 7095, CNRS - Universit\'e Pierre et Marie Curie (Univ Paris 06), 75014 Paris, France}
\affiliation{Department of Physics \& Astronomy, University of Waterloo, Waterloo, ON N2L 3G1, Canada}
\affiliation{Perimeter Institute for Theoretical Physics, Waterloo, ON N2L 2Y5, Canada}
\affiliation{Canadian Institute for Theoretical Astrophysics, Toronto, ON M5S 3H8, Canada}

\author{Michael S. Warren}
\affiliation{Theoretical Division, Los Alamos National Laboratory, Los Alamos, NM 87545, USA}

\date{\today}

\begin{abstract}
Galaxy bias, the unknown relationship between the clustering of galaxies and the underlying dark matter density field is a major hurdle for cosmological inference from large-scale structure. While traditional analyses focus on the absolute clustering amplitude of high-density regions mapped out by galaxy surveys, we propose a relative measurement that compares those to the underdense regions, cosmic voids. On the basis of realistic mock catalogs we demonstrate that cross correlating galaxies and voids opens up the possibility to calibrate galaxy bias and to define a static ruler thanks to the observable geometric nature of voids. We illustrate how the clustering of voids is related to mass compensation and show that volume-exclusion significantly reduces the degree of stochasticity in their spatial distribution. Extracting the spherically averaged distribution of galaxies inside voids from their cross correlations reveals a remarkable concordance with the mass-density profile of voids.
\end{abstract}


\maketitle

\textit{Introduction.}---The distribution of matter in the Universe plays a key role for our understanding of cosmology. Galaxy surveys provide the means to map out this cosmic large-scale structure (LSS) in three dimensions, furnishing a cornerstone of observational cosmology (e.g., \cite{EUCLID,SDSS,VIPERS}). This information is given in the form of galaxy locations and is typically condensed into a single function of scale, such as the galaxy power spectrum, from which cosmological parameters are inferred. However, galaxies are not the only footprint of LSS those surveys provide. In fact, they observe regions in the Universe containing hardly any galaxies at all, so-called cosmic voids. In this Letter we argue additional information can be gained from including voids into the statistical inference process of LSS, a technique that is already applicable to existing data~\cite{Sutter2012a}.

\textit{Void model.}---In analogy to the well studied halo model~\cite{Cooray2002}, we define a void model of LSS which similarly assumes the entire matter distribution to be described as a superposition of voids. In this manner, the cross-power spectrum between void centers and galaxies can be split into two terms as a function of wave number $k$ -- a one-void (or shot noise) term 
\begin{equation}
P_{\void\gal}^{(1\mathcal{V})}(k)=\frac{1}{\bar{n}_\void\bar{n}_\gal}\int \frac{\mathrm{d}n_\void(r_\void)}{\mathrm{d}r_\void}N_\gal(r_\void)\;u_\void(k|r_\void)\;\mathrm{d}r_\void \;, \label{P1void}
\end{equation}
which only considers correlations between the $N_\gal$ galaxies and the void center within any given void of radius $r_\void$, and a two-void term responsible for correlations between galaxies and void centers in distinct voids,
\begin{multline}
P_{\void\gal}^{(2\mathcal{V})}(k)=\frac{1}{\bar{n}_\void\bar{n}_\gal}\iint \frac{\mathrm{d}n_\void(r_\void)}{\mathrm{d}r_\void}\frac{\mathrm{d}n_\gal(m_\gal)}{\mathrm{d}m_\gal}b_\void(r_\void)b_\gal(m_\gal) \\
\times u_\void(k|r_\void)P_{\matter\matter}(k)\;\mathrm{d}r_\void\mathrm{d}m_\gal \;. \label{P2void}
\end{multline}
$n_\void$, $n_\gal$ are, respectively the number density functions of voids of radius $r_\void$, and of galaxies with host-halo mass $m_\gal$. Their corresponding linear bias parameters are $b_\void$ and $b_\gal$, we neglect higher-order bias terms. $u_\void(k|r_\void)$ describes the density profile for voids of radius $r_\void$ in Fourier space and $P_{\matter\matter}(k)$ the auto-power spectrum of dark matter~\footnote{$r_\void$ and $u_\void(k|r_\void)$ are to be understood as ensemble-averaged quantities, since individual voids are not spherical. We treat galaxies as pointlike, $u_\gal(k|m_\gal)=1$, as their finite size can be neglected on the scales of interest.}. For a narrow range in $r_\void$, the total void-galaxy cross-power spectrum becomes
\begin{equation}
P_{\void\gal}(k) \simeq b_\void b_\gal u_\void(k)P_{\matter\matter}(k) + \bar{n}_\void^{-1}u_\void(k)\;, \label{Pvg}
\end{equation}
where we dropped all explicit dependences on void radius and host-halo mass for simplicity. Analogously, for the auto-power spectra of voids and galaxies the model yields
\begin{eqnarray}
&P_{\void\void}(k) \simeq b_\void^2 u^2_\void(k)P_{\matter\matter}(k) + \bar{n}_\void^{-1} \;,& \label{Pvv} \\
&P_{\gal\gal}(k) \simeq b_\gal^2 P_{\matter\matter}(k) + \bar{n}_\gal^{-1} \;,& \label{Pgg}
\end{eqnarray}
so in the high sampling limit of $\bar{n}_\void^{-1}, \bar{n}_\gal^{-1} \ll P_{\matter\matter}$, the void density profile in Fourier space can be estimated as
\begin{equation}
u_\void(k) \simeq \frac{b_\gal P_{\void\gal}(k)}{b_\void P_{\gal\gal}(k)}\simeq\frac{P_{\void\gal}(k)}{P_{\gal\gal}(k)}\times\left.\frac{P_{\gal\gal}(k)}{P_{\void\gal}(k)}\right\vert_{k\to0} \;. \label{profile}
\end{equation}
Its relation to configuration space can be expressed via
\begin{equation}
u_\void(k)=\frac{\bar{\rho}}{\delta m}\int_0^\infty u_\void(r)\frac{\sin(kr)}{kr}4\pi r^2\;\mathrm{d}r \;, \label{Hankel}
\end{equation}
where $u_\void(r)=\rho_\void(r)/\bar{\rho}-1$ is the spherically averaged relative deviation of mass density in a void from the mean value $\bar{\rho}$ across the Universe and $\delta m$ the void's uncompensated mass. The profile is normalized such that $u_\void(k\to0)=1$~\cite{Cooray2002}, i.e.,
\begin{equation}
\delta m = \bar{\rho}\int_0^\infty u_\void(r)4\pi r^2\;\mathrm{d}r \;. \label{Norm}
\end{equation}
From Eq.~(\ref{Hankel}) we also have $u_\void(k\to\infty)=0$, assuming $|ru_\void(r)|<\infty$. Likewise, $u_\void(r\to\infty)=0$, as voids are local structures with a finite extent. The remaining limit is determined by the matter density in the void center, which for an empty void yields $u_\void(r\to0)=-1$.

\textit{Compensation.}---For the particular case of a compensated void, whose density decrement in its center is exactly balanced by an overdense wall around it, the normalization condition $u_\void(k\to0)=1$ cannot be enforced, as $\delta m=0$. Because of the geometric definition of voids it is more meaningful to normalize Eq.~(\ref{Hankel}) by the volume of a void region $V_\void$ (including its compensation wall), which yields a renormalized profile $b_\void(k)$ with
\begin{equation}
\frac{\delta m}{\bar{\rho}V_\void} u_\void(k) \equiv b_\void(k) \;, \label{b(k)}
\end{equation}
such that $|b_\void(k)|<\infty$ for all $\delta m$. This matches the large-scale clustering properties of voids in the linear regime to the nonlinear domain of the internal void structure, so it can be interpreted as a scale-dependent void bias $b_\void(k)\equiv b_\void u_\void(k)$. In particular, it agrees with the fact that compensated structures ($\delta m=0$) do not generate any large-scale power ($b_\void=0$), because they only rearrange mass locally~\cite{Cooray2002}. Equation~(\ref{b(k)}) gives a simple explanation of linear bias: it is the uncompensated mass $\delta m= m-\bar{\rho}V$ of a tracer compared to the mass of an equally sized region of volume $V$ of the background, $b=\delta m/\bar{\rho}V$. This is indeed predicted by so-called Poisson cluster models, where halo bias arises as a consequence of mass conservation, or, in other words: the distribution of halos depends on their environment~\cite{Sheth1998,Abbas2007}. Thanks to the symmetry of the initial Gaussian field, this argument applies to voids just as well, with the advantage of the volume of voids being observationally more accessible than the volume of halos. Unfortunately, uncompensated mass is not directly observable (except via gravitational lensing~\cite{Bolejko2013,Krause2013,Higuchi2013}), but we can define a similar relation to Eq.~(\ref{b(k)}) using galaxies as tracer particles, with the replacements $\delta m\to\delta N_\gal=N_\gal-\bar{n}_\gal V$ and $\bar{\rho}\to\bar{n}_\gal$. This yields the relative bias between voids and galaxies,
\begin{equation}
\frac{\delta N_\gal}{\bar{n}_\gal V_\void} u_\void(k) \simeq 
\frac{b_\void(k)}{b_\gal} \;. \label{relb(k)}
\end{equation}

\begin{figure*}[!t]
\centering
\resizebox{\hsize}{!}{
\includegraphics[trim=0 -8 0 0,clip]{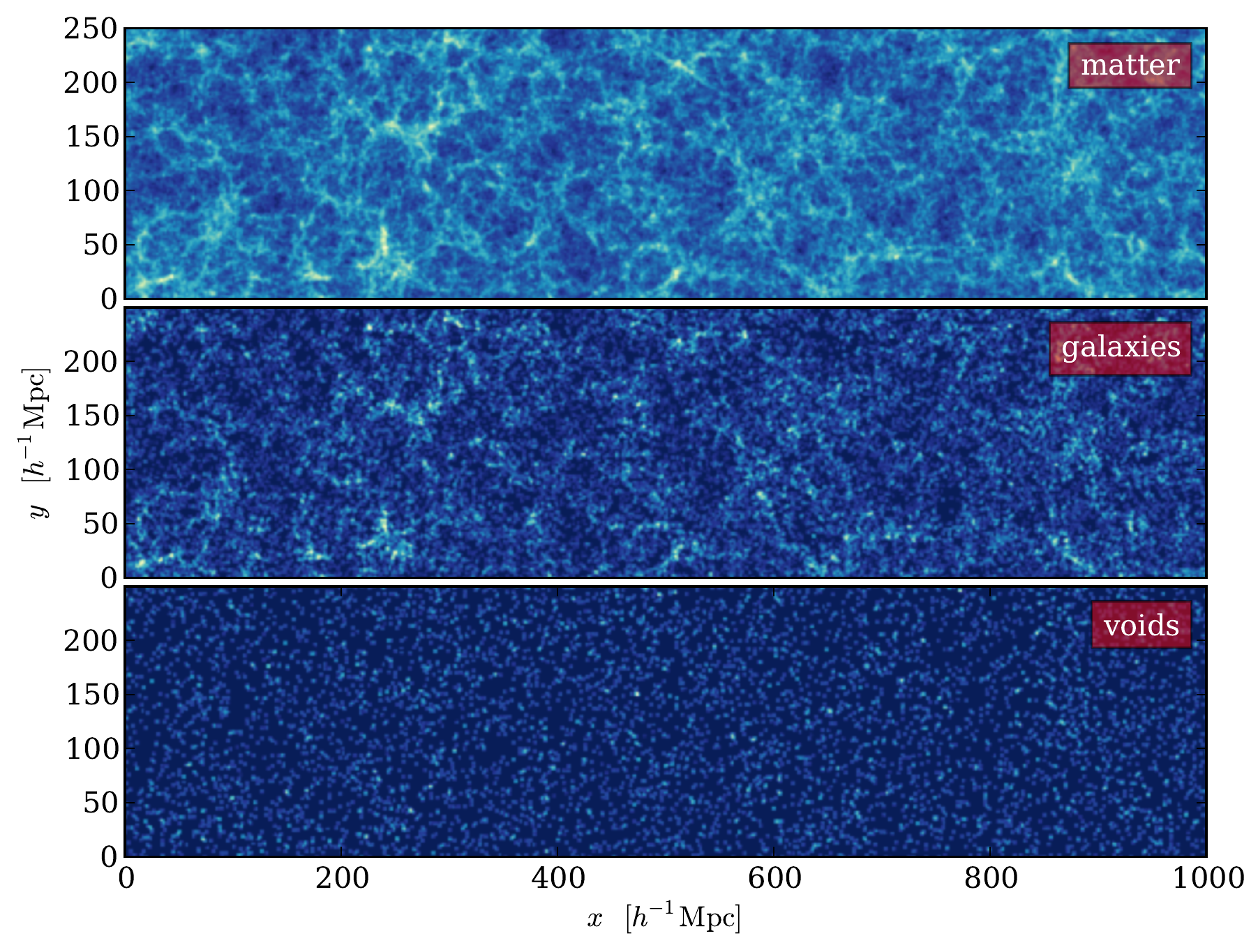}
\includegraphics[trim=0 0 0 0,clip]{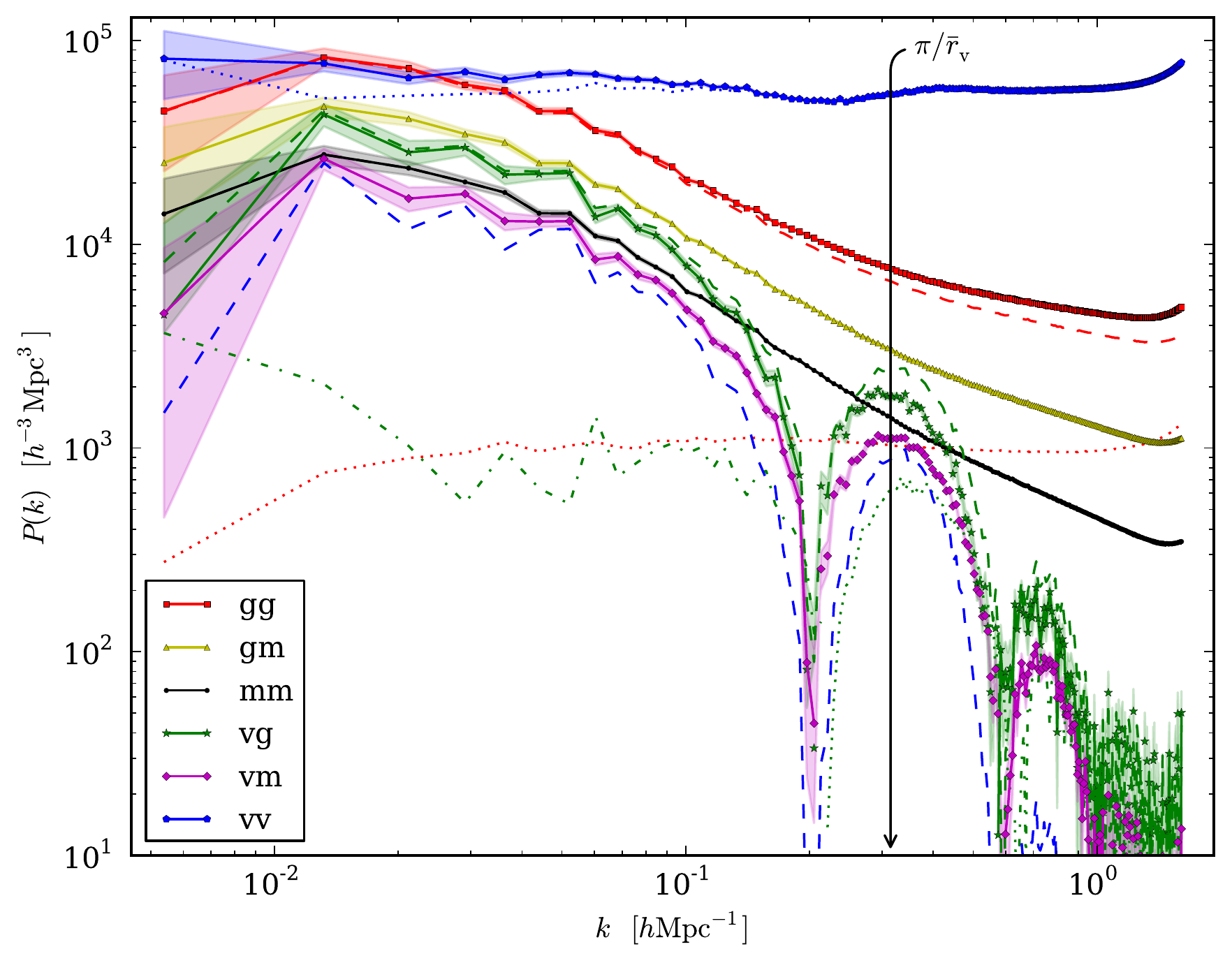}}
\caption{LEFT: Projected density fields of dark matter (top), galaxies (middle), and voids (bottom) in a $250\hMpc$ slice of the simulation box. RIGHT: Auto- and cross-power spectra for all possible combinations of dark matter, galaxies, and voids (solid lines connected by symbols, line omitted when negative). Subtracting out shot noise (drawn in dotted if positive and dot-dashed if negative) yields the dashed lines. Shaded bands show $1\sigma$ uncertainties estimated from scatter in the bin average.}
\label{fig1}
\end{figure*}

\textit{Simulations.}---For our numerical analysis we employ the 2HOT $N$-body code~\cite{Warren2013} to evolve $2048^3$ cold dark matter particles in a $1\hGpc$ box of a \textsc{planck} cosmology~\cite{Planck2013}. From the final snapshot at redshift $z=0$ we generate a halo catalog using the \textsc{rockstar} halo finder~\cite{Behroozi2013} with an overdensity threshold of $\delta\ge200$ to define virialized objects and utilize a standard halo occupation distribution model~\cite{Tinker2005,Zheng2007} with parameters adapted to the \textsc{sdss dr7}~\cite{Zehavi2011} to obtain a realistic mock galaxy sample. The resulting distribution of $\sim2\times10^7$ galaxies with host-halo masses $m_\gal\gtrsim2\times10^{11}\hMsun$ and a mean separation of $\sim3.7\hMpc$ is then used to generate a void catalog based on the \textsc{Zobov}~\cite{Neyrinck2008} code, which finds density minima in a Voronoi tessellation of the tracer particles and grows basins around them applying the watershed transform~\cite{Platen2007}. This gives rise to $\sim1\times10^5$ voids with effective radii $5\hMpc\lesssim r_\void\lesssim150\hMpc$~\footnote{We restrict ourselves to zones with minimum underdensity $\delta\le-0.8$ when merging them into voids and define void centers as averages of the void's particle positions, weighted by their Voronoi cell-volume; see~\cite{Lavaux2012,Sutter2012a}.}. Power spectra are obtained by Fourier transforming a cloud-in-cell interpolation of the tracer particles on a cubic mesh of $512^3$ grid points and subsequent shell averaging.

\textit{Clustering statistics.}---Figure~\ref{fig1} depicts slices of the density fields used in our analysis, going from dark matter over galaxies to voids. While galaxies trace out the most overdense structures of the dark matter, voids are distributed more evenly than the underlying LSS. Their distribution appears more noisy, as they are sparser than galaxies and dark matter particles. The corresponding power spectra are shown in the right panel of Fig.~\ref{fig1}, where we selected all galaxies with host-halo masses $m_\gal\ge1\times10^{13}\hMsun$ and voids of radii $9\hMpc\le r_\void\le 11\hMpc$ with an average $\bar{r}_\void\simeq10\hMpc$. While the auto-power spectrum of galaxies closely follows the shape of the underlying matter power spectrum, this is not the case for voids. Here the shot noise term in Eq.~(\ref{Pvv}) dominates over the bare clustering term. The latter is suppressed due to the low bias parameter of the selected voids, $b_\void\simeq0.8$.

The cross-power spectrum between galaxies and voids largely avoids this problem, as shot noise turns out to be much lower in this case. In fact, its magnitude remains below the bare clustering power everywhere. We can identify two regimes as suggested by Eq.~(\ref{Pvg}): linear clustering with constant bias on large scales and a nonlinear suppression of power on small scales due to the void profile. The cross-power spectrum even turns negative and reaches a minimum at $k\sim\pi/\bar{r}_\void$, a consequence of galaxy-void exclusion, which also causes the shot noise to be much lower than expected from Eq.~(\ref{Pvg})~\cite{Baldauf2013}. As a sanity check we also show cross-power spectra of each tracer with the dark matter. We find consistency with Eq.~(\ref{Pvg}) when setting galaxy bias to $1$ and neglecting shot noise due to the high density of dark matter particles.

\begin{figure*}[!t]
\centering
\resizebox{\hsize}{!}{
\includegraphics[trim=0 0 0 0,clip]{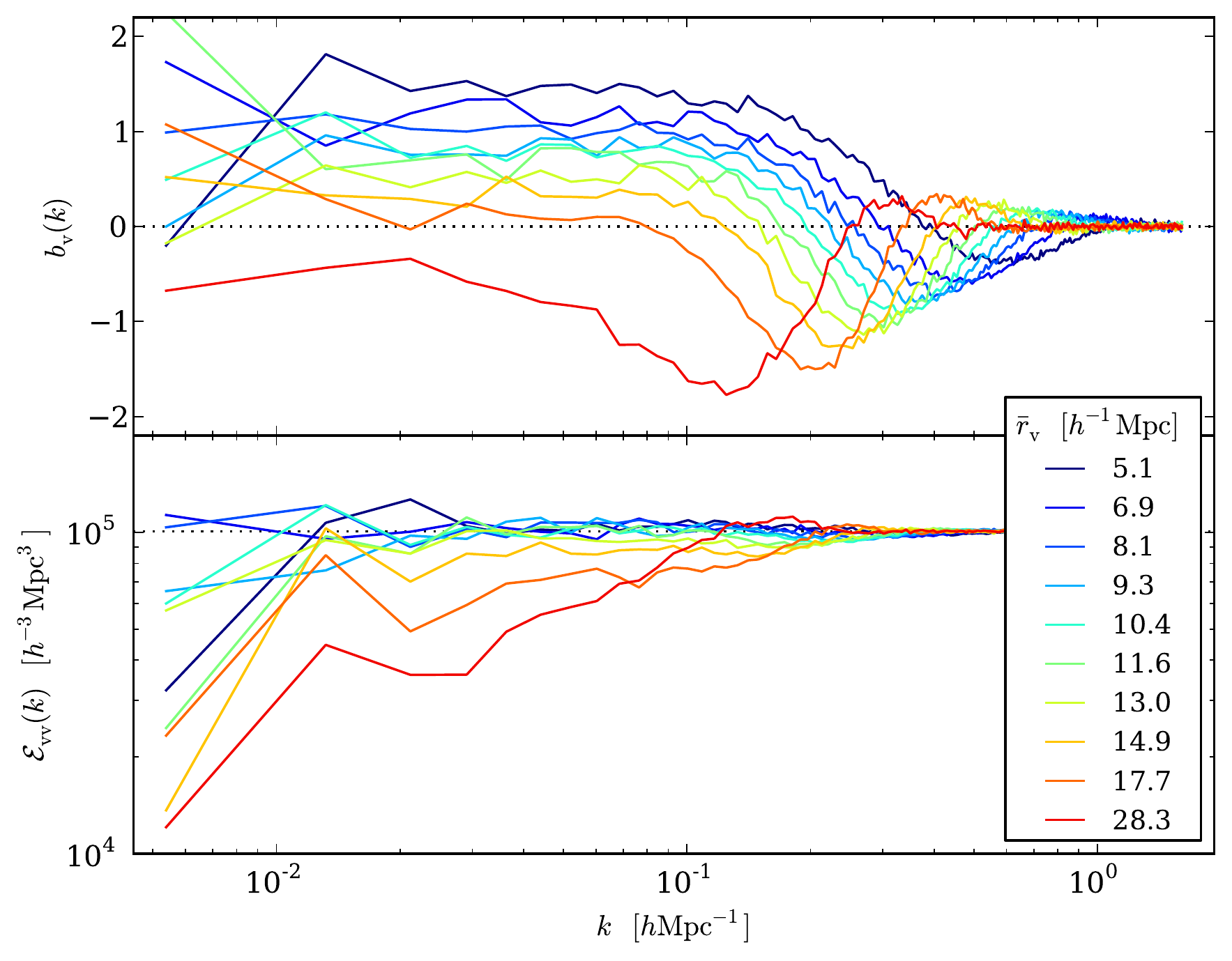}
\includegraphics[trim=0 -4 0 0,clip]{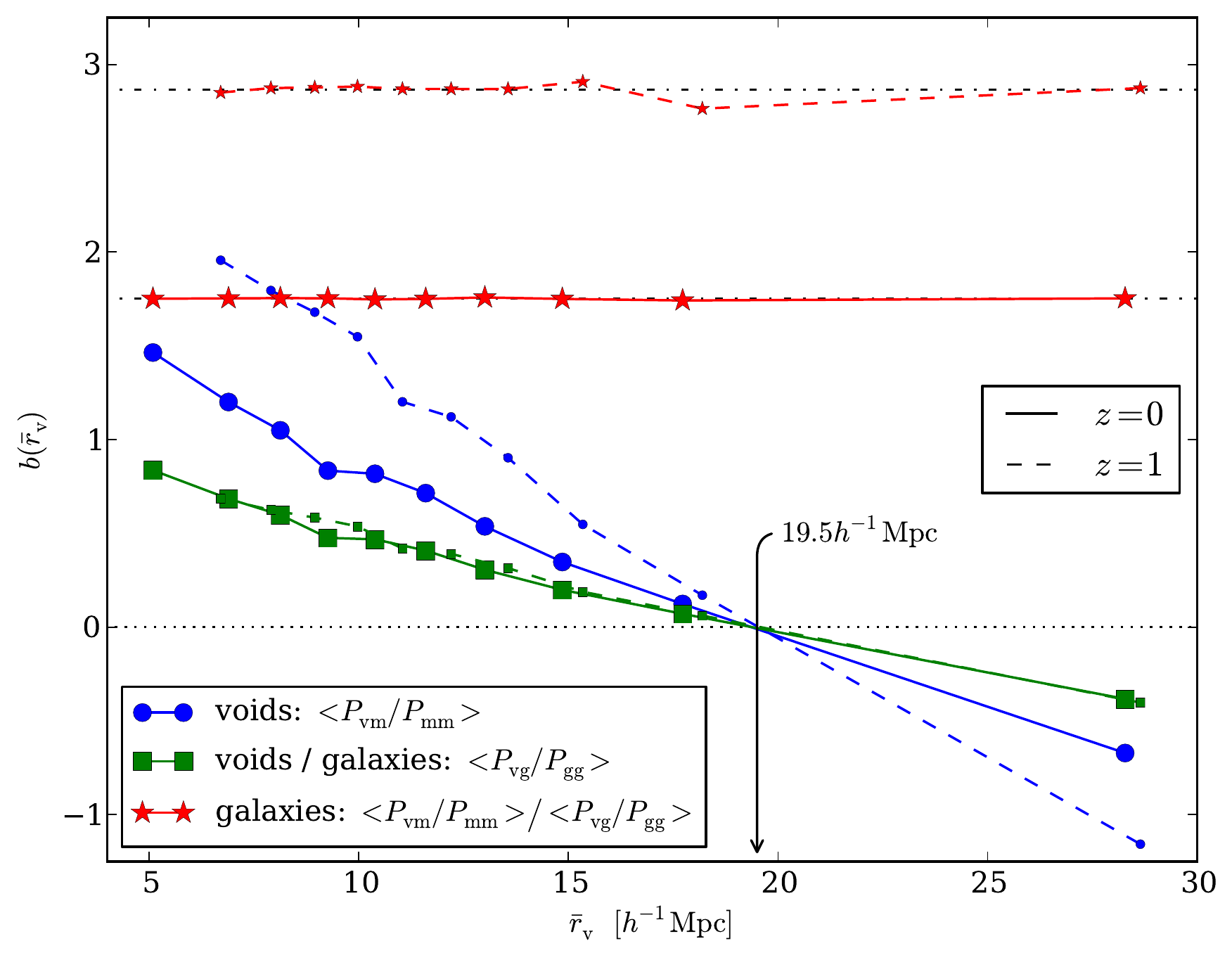}}
\caption{LEFT: Scale-dependent bias (top) and shot noise (bottom) of voids from $10$ equal number-density bins of increasing void radius with mean values shown in the inset. Dotted lines show reference values of $0$ (top) and $1/\bar{n}_\void$ (bottom). RIGHT: Large-scale bias of voids (circles) and relative void-galaxy bias with galaxies of host-halo mass $m_\gal\ge1\times10^{13}\hMsun$ (squares) as a function of bins in void radius $\bar{r}_\void$ (same as left). The bias of the galaxy sample (dot-dashed lines) is obtained via the ratio of the two (stars). Results are depicted for redshift $z=0$ (large symbols, solid lines) and $z=1$ (small symbols, dashed lines).}
\label{fig2}
\end{figure*}

Computing dark matter cross-power spectra allows us to determine the scale-dependent nonlinear bias $b_x(k)= P_{x\matter}/P_{\matter\matter}$ of a tracer $x$ and shot noise $\mathcal{E}_{xy}(k)= P_{xy}-P_{x\matter}P_{y\matter}/P_{\matter\matter}$ for tracers $x$ and $y$ directly from the simulations~\cite{Hamaus2010}. Figure~\ref{fig2} shows these two quantities for voids in $10$ consecutive bins in radius, each containing the same number of voids. On large scales, void bias is scale independent, as linear theory suggests. While small voids can have bias parameters larger than unity due to their occurrence in overdense structures such as filaments and sheets, large voids are increasingly antibiased tracers of the dark matter. This behavior can be described in terms of compensation as suggested by Eq.~(\ref{b(k)}): small (large) voids are overcompensated (undercompensated) with more (less) material in their surrounding than in their interior.
Exact compensation cannot lead to any large-scale power, so the linear bias parameter must vanish. Here this is the case for voids with radius $\bar{r}_\void\simeq20\hMpc$, which defines a compensation scale. Overcompensated voids will ultimately disappear under the gravitational evolution of their surrounding (void-in-cloud), while undercompensated ones will merge with other voids (void-in-void)~\cite{Sheth2004,Ceccarelli2013}. The compensation scale is not unique, but decreases when the density of tracers that define the voids is increased. However, thanks to the self-similar nature in the distribution of voids, it is uniquely determined at any given tracer density~\cite{Sutter2013}.

Towards smaller scales void bias becomes scale dependent, decreasing to a minimum located at the exclusion scale of the void sample, $k_\mathrm{exc}\sim\pi/\bar{r}_\void$. Any two voids of similar size do not overlap and are thus anticorrelated on this scale~\cite{Colberg2005,Padilla2005}. As $k$ increases further, $b_\void(k)$ crosses zero around $k\sim 2\pi/\bar{r}_\void$ to reach a local maximum and finally asymptotes to zero in the limit $k\to\infty$. Note that protohalos show a very similar behavior~\cite{Baldauf2013}, suggesting voids to preserve the properties of the initial conditions much better than halos or galaxies. The effect of exclusion is also visible in the void shot noise, as shown in the lower-left panel of Fig.~\ref{fig2}. While small voids tend to sample the density field as expected from Poisson statistics with shot noise given by $1/\bar{n}_\void$, larger voids can have substantially lower values. Because of their large size and high volume fraction in the Universe, exclusion effects are, hence, more important in the clustering statistics of voids as compared to halos or galaxies~\cite{Baldauf2013}.

Unfortunately, bias and shot noise of neither galaxies nor voids can be determined observationally, because the dark matter distribution is unknown. We therefore advocate the use of comparative quantities for any inference from LSS, such as the relative bias between galaxies and voids. In contrast to masses, volumes are observed directly and may thus be used to calibrate galaxy bias. An example is given in the right panel of Fig.~\ref{fig2}, where we plot $b_\void$ in void-radius bins taken from the left panel, averaged over all modes with $k\le0.05h{\rm Mpc}^{-1}$. Although not directly observable, $b_\void(\bar{r}_\void)$ can be calculated from theory or calibrated to simulations based on the measured geometry and abundance of voids in observations~\footnote{This requires assuming a specific cosmology, but the dependence on standard cosmological parameters in relative clustering amplitudes is rather weak compared to absolute ones, such as the power spectrum.}. When divided by the observed relative bias between voids and galaxies, $b_\gal$ can be determined from each void population. This does not depend on any prior assumptions on how we populated dark matter halos with galaxies, since we checked that all of our results are identical when defining voids in the distribution of halos instead of galaxies.

The zero crossing of the relative bias provides the compensation scale. As it coincides with the zero crossing of the void bias $b_\void(\bar{r}_\void)$, this suggests that if voids are compensated by galaxies ($\delta N_\gal=0$), they are also compensated in mass ($\delta m=0$) and vice versa. If mass conservation is assumed, only compensated voids should remain compensated in the course of cosmological evolution and may therefore serve as a static ruler on scales smaller than the baryon acoustic oscillations~\cite{Eisenstein2005} (conversely, mass conservation can be tested on cosmological scales if compensated voids are assumed to be static rulers). In contrast to a standard ruler the comoving size of a static ruler is not necessarily determined by a physical scale, but it is conserved and thus can be used to probe the expansion history of the Universe. Figure~\ref{fig2} suggests that compensated voids may potentially be used as static rulers, as the zero crossing of $b_\void(\bar{r}_\void)$ appears to remain at $\bar{r}_\void\simeq20\hMpc$ even at redshift $z=1$, whereas its slope and average void radii increase~\footnote{Although in realistic surveys the varying tracer number density with redshift also influences the compensation scale due to selection effects, one can account for that by subsampling the tracers according to their selection function.}. For the relative bias between voids and galaxies even the slope is not affected by a change in redshift, suggesting the evolution of bias to be canceled in this ratio.

\begin{figure}[!t]
\centering
\resizebox{\hsize}{!}{
\includegraphics[trim=0 0 0 0,clip]{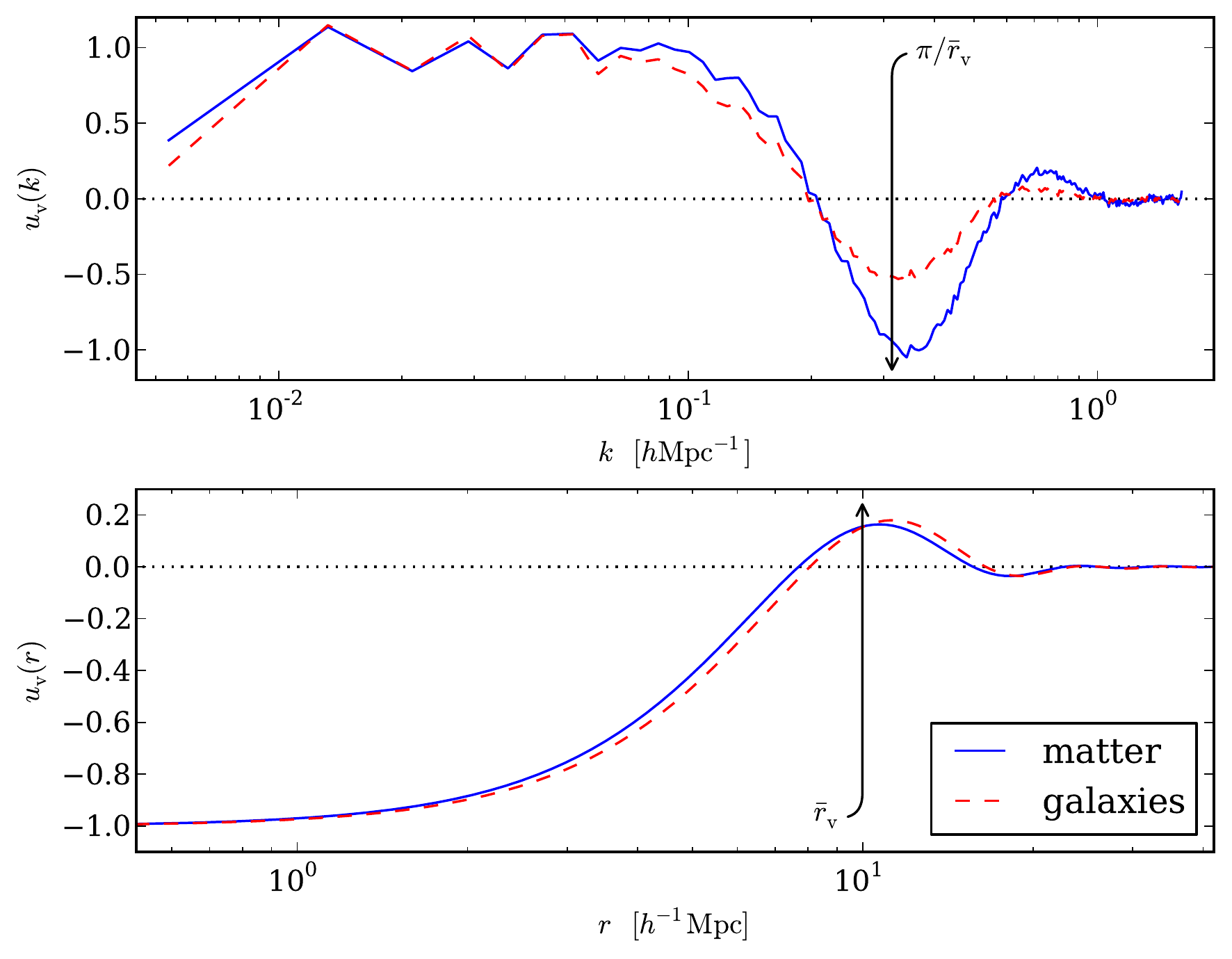}}
\caption{Density profile of $\bar{r}_\void=10\hMpc$ voids as traced by dark matter particles (solid) and by galaxies (dashed) in Fourier space (top) and in configuration space (bottom).}
\label{fig3}
\end{figure}

\textit{Void profile.}---According to Eq.~(\ref{profile}) the void density profile can be estimated from the ratio between void-galaxy and galaxy-galaxy power spectra, which effectively cancels out the two-void component of the void model (as shown in~\cite{Neyrinck2005} for halos). We test this approach by comparing it to the void profile as traced by dark matter particles, using $u_\void(k)\simeq P_{\void\matter}/b_\void P_{\matter\matter}$, which is not affected by shot noise. From the top panel in Fig.~\ref{fig3} it is evident that both agree very well with each other. The difference is caused by shot noise contributions in $P_{\void\gal}$ and $P_{\gal\gal}$: when subtracted, the two profiles are consistent with each other. An inverse Fourier transform yields the void profile in configuration space $u_\void(r)$~\footnote{In order to suppress nonlinearities we only integrated up to $k=2\pi/\bar{r}_\void$, which is equivalent to a top-hat smoothing of that scale in Fourier space. Consequently, we normalize the profile to be empty in the center, i.e., $u_\void(r\to0)=-1$.}, shown in the lower panel of Fig.~\ref{fig3}. The profiles traced by mock galaxies and dark matter show a remarkable agreement, both exhibiting a steep compensation wall with a peak at $\bar{r}_\void\simeq10\hMpc$. Similar profile shapes have been obtained from stacked voids found in simulations~\cite{Lavaux2012} and observational data~\cite{Ceccarelli2013,Pisani2013,Sutter2012a}, a fact that consolidates the presented framework of the void model.

\textit{Outlook.}---
Several applications could potentially benefit from the presented analysis. One example is the Alcock-Paczynski test~\cite{Alcock1979}, which has already been applied to stacked density profiles of voids from both simulations~\cite{Lavaux2012} and observations~\cite{Sutter2012b}. An alternative analysis can be conducted with the two-dimensional power spectrum $P_{\void\gal}(k,\mu)$ in redshift space, where $\mu$ is the cosine of the angle between a Fourier mode and the line of sight. Here, the exclusion scale $k\sim\pi/\bar{r}_\void$ results in a high-contrast ringlike structure, distorted by peculiar motions of galaxies along the line of sight~\cite{Paz2013}. We leave the investigation of redshift-space distortions for further study, as their impact on the clustering statistics of voids is rather weak~\cite{Padilla2005,Lavaux2012}.

The ratio $P_{\void\gal}(k)/P_{\gal\gal}(k)$ may serve as an ideal estimator for all types of primordial non-Gaussianity that induce scale-dependent corrections to the linear bias~\cite{Dalal2008}. The signal on such corrections increases with the range in linear bias that can be probed with multiple tracers~\cite{Hamaus2011} and can thus be enhanced by including even antibiased objects in the analysis. 
Further accuracy may be obtained by utilizing optimal weights for voids in order to suppress their stochasticity, as  demonstrated for halos in~\cite{Seljak2009b,Hamaus2010}. The same techniques can be applied in redshift space to put constraints on the growth rate of structure formation and to test general relativity on cosmological scales~\cite{Hamaus2012,Yoo2012}.

Finally, the detailed shape and the compensation of void density profiles can be used to probe dark energy~\cite{Bos2012} and modified gravity models~\cite{Li2011,Belikov2013,Clampitt2013}. In contrast to the void number function, which strongly depends on the type of tracer particle to define a void~\cite{Jennings2013}, void density profiles extracted from correlations with either dark matter or galaxies are fairly consistent with each other. Moreover, the influence of nonlinear corrections in the clustering of voids is expected to be milder as compared to galaxies or dark matter~\cite{Neyrinck2013,Leclercq2013}. We plan to report a number of results along these lines in the near future, including applications to observational data.

\begin{acknowledgments}
We thank St\'ephane Colombi, Neal Dalal, Vincent Desjacques, Jens Jasche, Florent Leclercq, Alice Pisani, Joe Silk, and Douglas Spolyar for helpful discussions and Jeremy Tinker for providing his HOD code. We acknowledge support of the LANL Institutional Computing program for the simulations presented here. This work was partially supported by NSF Grant No.~AST-0908902. B.D.W. is partially supported by a senior Excellence Chair by the Agence Nationale de Recherche (ANR-10-CEXC-004-01) and a Chaire Internationale at the Universit\'e Pierre et Marie Curie. G.L. acknowledges support from CITA National Fellowship and financial support from the Government of Canada Post-Doctoral Research Fellowship. Research at Perimeter Institute is supported by the Government of Canada through Industry Canada and by the Province of Ontario through the Ministry of Research and Innovation.
\end{acknowledgments}

\bibliography{ms.bib}
\bibliographystyle{apsrev.bst}

\end{document}